\documentclass[aps,physrev,reprint,superscriptaddress,nofootinbib,amsmath,amssymb,a4paper,showkeys]{revtex4-2}
\usepackage[T1]{fontenc}
\usepackage{newpxtext}                     % Palatino body text
\usepackage{eulervm} % Euler math font, clean and upright
\usepackage{mathtools}
\DeclarePairedDelimiter\bra{\langle}{\rvert}
\DeclarePairedDelimiter\ket{\lvert}{\rangle}
\DeclarePairedDelimiterX\braket[2]{\langle}{\rangle}{#1 \delimsize\vert #2}
\DeclarePairedDelimiterX\matrixel[3]{\langle}{\rangle}{#1 \delimsize\vert #2 \delimsize\vert #3}
\DeclarePairedDelimiterX\comm[2]{[}{]}{#1, #2}
\DeclarePairedDelimiterX\acomm[2]{\{}{\}}{#1, #2}
\DeclarePairedDelimiter\expect{\langle}{\rangle}

\usepackage{bm}                            % Bold math variables
\usepackage{siunitx}                       % Modern usage: \qty{\sim 200}{\kelvin}
\usepackage{derivative}
\usepackage{mleftright}
\mleftright

\usepackage{graphicx}
\usepackage{dcolumn}                       % Align table columns on decimal point
\newcolumntype{d}[1]{D{.}{\cdot}{#1}}
\usepackage{tikz}
\usetikzlibrary{decorations.markings}
\usetikzlibrary{arrows.meta}               % for Slater-diagram electron arrows
\usepackage{xcolor,soul}                   % Text highlighting

\begin{document}

%\title{Working title: The great fractal strawberry}
\title{Self-attention is a Dyson equation for many-body machine learning}
\date{\today} % very useful for keeping track of hard-copy drafts

\author{Jarvist Moore Frost}
\affiliation{Department of Chemistry, Imperial College London, Exhibition Road, London SW7 2AZ, UK}
\affiliation{Department of Physics, Imperial College London, Exhibition Road, London SW7 2AZ, UK}
\email{jarvist.frost@imperial.ac.uk}


\begin{abstract}
Why does symmetric sum-pooling (DeepSets) predict ground-state potentials to sub-meV accuracy, yet fail for correlated wavefunctions and excited states? 
Sum-pooling is a mean-field (Hartree) model: its $\mathcal{O}(1/N)$ off-diagonal Jacobian suffices for nearsighted ground-state energies, but erases nonlocal two-body correlations. 
We show that self-attention is a real-space Dyson iteration whose Boltzmann (softmax) kernel acts as a two-body self-energy. 
Because physical Hamiltonians act pairwise, L\"owdin's 2-RDM theorem and the Galitskii--Migdal relation guarantee that iterating a two-body propagator captures the full energetics of correlated matter. 
When fitting a fractal backflow determinant, DeepSets blurs the nodal surface onto a mean-field envelope, whereas self-attention recovers the fractal attractor. 
Seen through this lens, the unreasonable effectiveness of attention is not an accident; in a universe built from pairwise interactions, two-body attention is all you need.
\end{abstract}

\maketitle

\section{Introduction}

Neural network wavefunction Ans\"atze often feed electron-pair separations directly into the network\cite{Hermann2023}, a brute-force $\mathcal{O}(N^2)$ approach that bloats the architecture, while raw single-particle coordinates lose translational symmetry unless embedded periodically\cite{Cassella2023}. 
Symmetric sum-pooling (often via equivariant DeepSets) scalarises interactions between particles into one-body terms. 
By the universal approximation theorem, this is formally sufficient to encode everything about the system, but with finite capacity it is not tractable to extract many-body correlations from a single pooled vector. 
Yet in machine learning interatomic potentials for a system size of $N$ atoms, finite-cutoff equivariant message passing followed by DeepSets sum-pooling over local atomic energies $E_i$ ($E=\sum_{i=1}^N{E_i}$) achieves sub-meV\cite{Batatia2022} accuracy; whereas \textsc{PsiFormer}\cite{vonglehn2023selfattentionansatzabinitioquantum} replaces the pair features in \textsc{FermiNet}\cite{Pfau2020} with more simple multi-head self-attention over single-particle features, and achieves superior electronic correlation in molecular dimers. 

Attention\cite{Vaswani2017Attention} now underpins almost all areas of machine learning, improving performance while simplifying complex architectures. 
Conventionally framed in computer-science database terminology, the mechanism pairs queries with keys and normalises their dot products via a softmax to perform a ``fuzzy'' lookup over values. 
Following the example of atom-density representations written in Dirac notation\cite{Willatt2019DiracNotation}, we recast this database abstraction into the operator language of many-body physics.
At its core, self-attention applies learned linear projection matrices $W_{Q,K,V}$ to map single-particle states $\ket{h_i}$ into queries $\ket{q_i}=W_Q\ket{h_i}$, keys $\ket{k_j}=W_K\ket{h_j}$ and values $\ket{v_j}=W_V\ket{h_j}$, before normalising their query--key inner products with a Boltzmann (softmax) factor. 
This reduces to a state-dependent two-body propagator $\sum_{j=1}^N Z_i^{-1}\exp(\beta\braket{q_i}{k_j})\ket{v_j}$, normalised by the local partition function $Z_i = \sum_{n=1}^N \exp(\beta\braket{q_i}{k_n})$.

In this paper we show that DeepSets (scalar) pooling is a one-body mean-field (Hartree) description, explaining why it accurately captures additive, locally nearsighted quantities such as ground-state energies and forces. 
Remarkably, self-attention maps directly onto Dyson resummation of the nonlocal two-body propagator. 
Because physical Hamiltonians are governed by pairwise interactions and exact energetics reduce to two-body correlators\cite{Lowdin1955-2RDM}, iterative self-attention is formally sufficient to describe the one- and two-body observables (and hence exact energetics) of pairwise (physical, Coulomb) Hamiltonians.

Consequently, machine learning architectures directly targeting nonlocal quantum mechanical observables require self-attention rather than scalar sum-pooling. 
This equivalence to foundational completeness theorems of quantum matter explains why, across so many physical systems, ``attention is all you need''\cite{Vaswani2017Attention}. 
Here we make this connection between machine-learning architectures and many-body theory explicit, bridging the disparate notation and terminology of the two communities.

\section{Theory}

Consider a system of $N$ quantum particles or atoms at collective coordinates $\bm{R} = (\bm{r}_1, \dots, \bm{r}_N)$ with $\bm{r}_i \in \mathbb{R}^D$ in $D$ spatial dimensions, represented at layer $l$ (of $L$) of a machine-learning model by the set of permutation-equivariant single-particle states $h^{(l)} = \{\ket*{h_i^{(l)}}\}_{i=1}^N$. 
Both DeepSets and self-attention update operations can be written as a discrete two-body Dyson equation,
\begin{equation}
    \ket*{h_i^{(l+1)}} = \ket*{h_i^{(l)}} + \sum_{j=1}^N \Sigma_{ij}\!\left[h^{(l)}\right] \ket*{v_j^{(l)}},
    \label{eq:master_dyson}
\end{equation}
where $\ket*{v_j^{(l)}}$ is the single-particle value (message) sent by particle $j$ and $\Sigma_{ij}[h^{(l)}]$ is the \textit{scalar} effective two-body self-energy kernel coupling particle $i$ to particle $j$ as a functional of the effective layer state $h^{(l)}$.

%Layers $l$ within the machine learning model directly map to the level of recurrence in the Dyson equation; the model depth $L$ corresponds directly to the order of the Dyson iteration. 
Each layer $l$ in the machine-learning network performs one step of Dyson iteration; the dotal depth $L$ sets the order of resummation.

\subsection{Scalar DeepSets is the mean-field (Hartree) approximation}

In DeepSets\cite{Zaheer2017Deepsets}, interactions are aggregated via an unweighted symmetric sum (sum pooling) over all particles,
\begin{equation}
    \ket*{h_i^{(l+1)}} = \rho\!\left(\ket*{h_i^{(l)}}, \, \ket*{\bar{v}^{(l)}}\right), \qquad 
    \ket*{\bar{v}^{(l)}} = \frac{1}{N}\sum_{j=1}^N \phi\!\left(\ket*{h_j^{(l)}}\right),
    \label{eq:deepsets}
\end{equation}
where $\phi$ and $\rho$ are learned single-particle embedding and update functions (multilayer perceptrons), and $\ket*{\bar{v}^{(l)}}$ is the globally pooled value state. 
Setting individual particle values $\ket*{v_j^{(l)}} = \phi(\ket*{h_j^{(l)}})$ and comparing directly with Eqn.~\eqref{eq:master_dyson} reveals the physical loss: the pooled state $\ket*{\bar{v}^{(l)}}$  carries no target index $i$. 
Every particle experiences an identical background potential, corresponding to a uniform mean-field (Hartree) kernel $\overline{\Sigma}_{ij} = 1/N$. 
Consequently, the off-diagonal Jacobian measuring how particle $i$'s dressed state responds to particle $j$'s position scales as
\begin{equation}
    \overline{\bm{J}}_{ij} = \pdv*{h_i^{(l+1)}}{\bm{r}_j} = \frac{1}{N}\pdv*{\rho_i}{\bar{v}^{(l)}}\pdv*{v_j^{(l)}}{\bm{r}_j} \sim \mathcal{O}\!\left(\frac{1}{N}\right) \quad (i \neq j),
    \label{eq:jacobian_ds}
\end{equation}
%
% Don't include the full derivation, but might be useful? Could stick it in the SI. 
where $\rho_i \equiv \rho(\ket*{h_i^{(l)}}, \ket*{\bar{v}^{(l)}})$. 
When these pooled states are used downstream, specific pairwise correlations are diluted by $1/N$; worse, because Eqn.~\eqref{eq:jacobian_ds} factorises into an $i$-only factor $\pdv*{\rho_i}{\bar{v}^{(l)}}$  times a $j$-only factor $\pdv*{v_j^{(l)}}{\bm{r}_j}$ independent of pair separatios $r_{ij} = \lvert\bm{r}_i - \bm{r}_j\rvert$, any steep local gradient at $\bm{r}_j$ broadcasts spurious crosstalk to every distant spectator $i$. 

Physically this pathology means that a close approach between two particles jogs all the other particles in the system, as the large local interaction has to be modelled by the mean-field. 
As with any mean-field theory, this cannot support strong correlation. 

% HOLD MY BEER
% Mattuck-style diagram, generated with Google Gemini Argon
\begin{figure}[t]
\centering
\resizebox{\columnwidth}{!}{%
\begin{tikzpicture}[
    >=Stealth,
    font=\small,
    bare/.style={
        line width=0.65pt,
        postaction={decorate},
        decoration={markings, mark=at position 0.56 with {\arrow[line width=0.65pt]{Stealth[length=4.8pt, width=3.8pt]}}}
    },
    dressed/.style={
        double,
        double distance=1.5pt,
        line width=0.55pt,
        postaction={decorate},
        decoration={markings, mark=at position 0.58 with {\arrow[line width=0.65pt, double=none]{Stealth[length=5.4pt, width=4.5pt]}}}
    },
    loopdressed/.style={
        double,
        double distance=1.5pt,
        line width=0.55pt,
        postaction={decorate},
        decoration={markings, mark=at position 0.25 with {\arrow[line width=0.65pt, double=none]{Stealth[length=5.0pt, width=4.2pt]}}}
    },
    intline/.style={
        line width=0.7pt,
        dashed
    },
    vertex/.style={
        circle,
        fill=black,
        inner sep=1.5pt
    }
]

% ============================================================
% PANEL (a): DeepSets = Hartree Tadpole (Mean-Field)
% ============================================================
\node[anchor=west, font=\bfseries\small] at (0.0, 2.55) {(a) DeepSets: Hartree mean-field tadpole ($\overline{\Sigma}_{ij} = 1/N$)};

% Dressed LHS state |h_i^{(l+1)}> (double line)
\node[anchor=east] at (0.88, 1.20) {$\ket*{h_i^{(l+1)}}$};
\draw[dressed] (1.00, 0.50) -- (1.00, 1.90);
\node[anchor=south, font=\footnotesize] at (1.00, 1.92) {$i$};
\node[anchor=north, font=\footnotesize] at (1.00, 0.48) {$i$};

% Equals sign
\node[font=\large] at (1.55, 1.20) {$=$};

% Residual skip connection |h_i^{(l)}> (unscattered single line G_0)
\draw[bare] (2.10, 0.50) -- (2.10, 1.90);
\node[anchor=south, font=\footnotesize] at (2.10, 1.92) {$i$};
\node[anchor=north, font=\footnotesize] at (2.10, 0.48) {$i$};
\node[anchor=west] at (2.22, 1.20) {$\ket*{h_i^{(l)}}$};

% Plus sign
\node[font=\large] at (3.65, 1.20) {$+$};

% Target line i: enters bare (single), exits dressed (double) above vertex \rho_i
\draw[bare] (4.85, 0.50) -- (4.85, 1.20);
\draw[dressed] (4.85, 1.20) -- (4.85, 1.90);
\node[anchor=south, font=\footnotesize] at (4.85, 1.92) {$i$};
\node[anchor=north, font=\footnotesize] at (4.85, 0.48) {$i$};
\node[vertex] at (4.85, 1.20) {};
\node[anchor=east, font=\footnotesize] at (4.72, 1.20) {$\rho_i$};

% 1/N interaction line to closed Hartree tadpole loop with vertex \phi_j
\draw[intline] (4.85, 1.20) -- node[above=1pt, font=\scriptsize] {$1/N$} (5.95, 1.20);
\node[vertex] at (5.95, 1.20) {};
\node[anchor=north east, font=\scriptsize] at (5.95, 1.15) {$\phi_j$};

% Closed Hartree loop (combined mean-field state |\bar{v}^{(l)}>, double line)
\draw[loopdressed] (6.40, 1.20) circle (0.45cm);
\node[font=\footnotesize] at (6.40, 1.20) {$j$};
\node[anchor=west, font=\small] at (6.95, 1.20) {$\ket*{\bar{v}^{(l)}}$};

% ============================================================
% PANEL (b): Self-Attention = Nonlocal Dyson Iteration
% ============================================================
\node[anchor=west, font=\bfseries\small] at (0.0, -0.35) {(b) Self-attention: nonlocal Dyson iteration ($\Sigma_{ij}^{\text{Attn}}[h^{(l)}]$)};

% Dressed LHS state |h_i^{(l+1)}> (double line)
\node[anchor=east] at (0.88, -1.70) {$\ket*{h_i^{(l+1)}}$};
\draw[dressed] (1.00, -2.40) -- (1.00, -1.00);
\node[anchor=south, font=\footnotesize] at (1.00, -0.98) {$i$};
\node[anchor=north, font=\footnotesize] at (1.00, -2.42) {$i$};

% Equals sign
\node[font=\large] at (1.55, -1.70) {$=$};

% Residual skip connection G_0 (unscattered single line)
\draw[bare] (2.10, -2.40) -- (2.10, -1.00);
\node[anchor=south, font=\footnotesize] at (2.10, -0.98) {$i$};
\node[anchor=north, font=\footnotesize] at (2.10, -2.42) {$i$};
\node[anchor=west] at (2.22, -1.70) {$\ket*{h_i^{(l)}}$};

% Plus summation over j (clean clearance before <q_i^{(l)}|)
\node[font=\small] at (3.65, -1.70) {$+\;\displaystyle\sum_{j=1}^N$};

% Target line i: enters bare (single), exits dressed (double) above vertex <q_i^{(l)}|
\draw[bare] (5.55, -2.40) -- (5.55, -1.70);
\draw[dressed] (5.55, -1.70) -- (5.55, -1.00);
\node[anchor=south, font=\footnotesize] at (5.55, -0.98) {$i$};
\node[anchor=north, font=\footnotesize] at (5.55, -2.42) {$i$};
\node[vertex] at (5.55, -1.70) {};
\node[anchor=east, font=\scriptsize] at (5.42, -1.70) {$\bra*{q_i^{(l)}}$};

% Directed Boltzmann self-energy kernel line (2.0 cm span: 5.55 <- 7.55)
\draw[intline, ->] (7.47, -1.70) --
    node[above=2pt, font=\scriptsize] {$\Sigma_{ij}^{\text{Attn}}[h^{(l)}]$}
    node[below=2pt, font=\scriptsize] {$Z_i^{-1}\mathrm{e}^{\beta\braket*{q_i}{k_j}}$}
    (5.63, -1.70);

% Partner j dressed value line |v_j^{(l)}> (double line) terminating at Key vertex |k_j^{(l)}>
\draw[dressed] (7.55, -2.40) -- (7.55, -1.70);
\node[anchor=north, font=\footnotesize] at (7.55, -2.42) {$j$};
\node[vertex] at (7.55, -1.70) {};
\node[anchor=west, font=\scriptsize] at (7.65, -1.55) {$\ket*{k_j^{(l)}}$};
\node[anchor=west, font=\scriptsize] at (7.65, -2.10) {$\ket*{v_j^{(l)}}$};

\end{tikzpicture}%
}

\caption{Following Mattuck\cite{mattuckGuideFeynmanDiagrams1992}, we express the two-body Dyson equation iteration as a Feynman diagram. 
\textbf{(a)} In DeepSets [Eqn.~\eqref{eq:deepsets}], summing $\phi(\ket*{h_j^{(l)}})$ inside a closed Hartree tadpole loop forms a mean-field average $\ket*{\bar{v}^{(l)}}$ that dresses particle $i$ via readout vertex $\rho_i$ with uniform weight $1/N$. 
Vertex operators $\rho$ and $\phi$ are multilayer perceptions (MLPs).
\textbf{(b)} In self-attention [Eqn.~\eqref{eq:attention_dyson}], the residual skip stream acts as the unscattered propagator $G_0$ (single line), while linear projections (vertex operators) $\bra*{q_i^{(l)}}$ and $\ket*{k_j^{(l)}}$ form a directed Boltzmann self-energy $\Sigma_{ij}^{\text{Attn}}[h^{(l)}]$ (dashed line) that combines partner $j$'s dressed value stream $\ket*{v_j^{(l)}}$ (double line) into particle $i$ with $\mathcal{O}(1)$ weight, dressing the outgoing propagator [Eqn.~\eqref{eq:dyson_gm}].}
\label{fig:mattuck_dyson}
\end{figure}
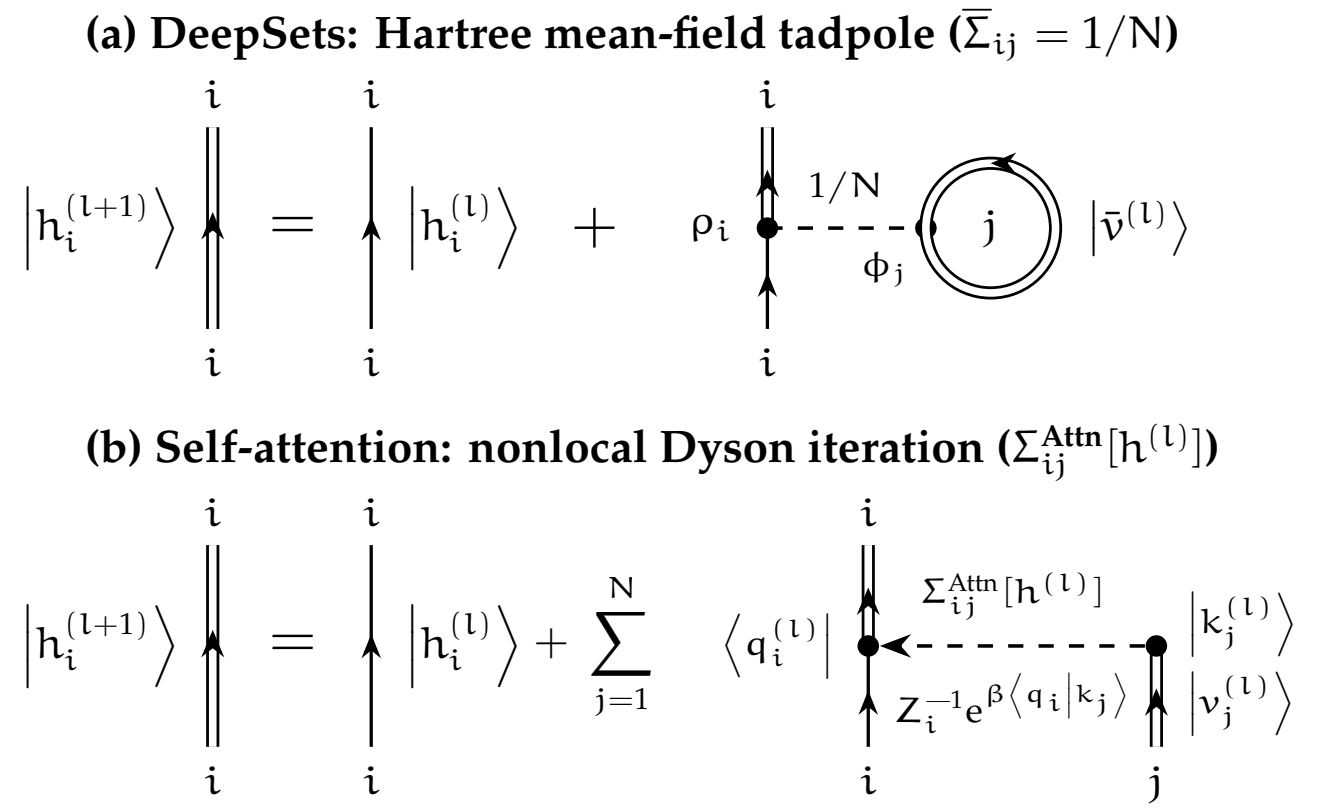% MATTUCK

\subsection{Self-attention is a Dyson equation}

Self-attention\cite{Vaswani2017Attention} replaces the uniform $\overline{\Sigma}_{ij} = 1/N$  weight in Eqn.~\eqref{eq:deepsets} with a state-dependent Boltzmann kernel.
Written in the motivating form of Dirac bra--ket notation, learned linear operators 
$W_Q, W_K \in \mathbb{R}^{d_k \times d_{\text{model}}}$ and $W_V \in \mathbb{R}^{d_{\text{model}} \times d_{\text{model}}}$ project each 
$d_{\text{model}}$-dimensional state $\ket*{h_i^{(l)}}$ into query 
$\ket*{q_i^{(l)}} = W_Q\ket*{h_i^{(l)}}$, key $\ket*{k_j^{(l)}} = W_K\ket*{h_j^{(l)}}$ (both of feature dimension $d_k$), and value 
$\ket*{v_j^{(l)}} = W_V\ket*{h_j^{(l)}}$ (model dimension $d_{\text{model}}$) states.
The update operation is then
\begin{equation}
    \ket*{h_i^{(l+1)}} = \ket*{h_i^{(l)}} + \sum_{j=1}^N \underbrace{\frac{\exp\!\left(\beta \braket*{q_i^{(l)}}{k_j^{(l)}}\right)}{Z_i^{(l)}}}_{\Sigma_{ij}^{\text{Attn}}[h^{(l)}]} \ket*{v_j^{(l)}},
    \label{eq:attention_dyson}
\end{equation}
where $Z_i^{(l)} = \sum_{n=1}^N \exp(\beta\braket*{q_i^{(l)}}{k_n^{(l)}})$ is the local canonical partition function at inverse effective temperature $\beta = d_k^{-1/2}$. 
This immediately generalises to $H$ parallel interaction channels in multi-head attention. 

In the high-temperature limit $\beta \to 0$, $\Sigma_{ij}^{\text{Attn}}[h^{(l)}] \to 1/N$ and Eqn.~\eqref{eq:attention_dyson} collapses to the DeepSets mean-field sum of Eqn.~\eqref{eq:deepsets}; whereas $\beta \to \infty$ freezes into a nearest-neighbour graph. 
The standard scaling $\beta = d_k^{-1/2}$ keeps $\beta\braket*{q_i^{(l)}}{k_j^{(l)}} \sim \mathcal{O}(1)$ in the correlated regime between mean-field melting and glassy freezing\cite{Ramsauer2021Hopfield}.

Here the inner product $\braket*{q_i^{(l)}}{k_j^{(l)}} = \matrixel*{h_i^{(l)}}{W_Q^\dagger W_K}{h_j^{(l)}}$ (where $W_Q^\dagger$ denotes the adjoint of $W_Q$)  acts as a learned, \textit{generally non-symmetric} ($W_Q \neq W_K$) two-body interaction vertex that dynamically reconstructs translationally invariant pair kernels (such as Gaussian radial basis or periodic plane waves) from single-particle states without storing explicit $\bm{r}_i - \bm{r}_j$ pair vectors.
% I replied to Christoph's email as well as I could, trying to justify my suggestion to add a PsiFormer style readout to X-MACE, but sent it fearing it was too handwavey for a mathmo---his emails always cut straight through my nonsense---and was then left home-alone quite literally holding the baby.
% I closed my laptop and brought her to the kitchen, cold sunlight streaming in, and tried to play with her on the mat, yet I couldn't stop thinking about why Attention actually worked, and how those little Attention tugs felt almost like Green's Functions, and the weight matrices just linear operators; there I was kneeling slack-jawed on the floor, watching Mattuck-style cartoons cascade down the wall like Tetris pieces, Hedin's pentagram creeping under the table, muttering under my breath about Dyson. 
% My four-month-old looked up at me wide-eyed and silent, wondering if this was what mother had been warning her about.
% Eid Ma Clack Shaw!
\footnote{
Constructing open-boundary one-body features $\ket*{h_i} = (\bm{r}_i, r_i^2, 1)^\top$ with $\bm{r}_i \in \mathbb{R}^D$ and $r_i = \lvert\bm{r}_i\rvert$, a symmetric vertex ($W_Q = W_K$) cannot couple $1_i$ to $r_j^2$ without introducing a translation-breaking $r_i^2 r_j^2$ cross-term. 
Non-symmetric $(D+1)\times(D+2)$ block weights 
$W_Q = \bigl(\begin{smallmatrix} 2\bm{I}_D & \bm{0} & \bm{0} \\ \bm{0}^\top & 0 & 1 \end{smallmatrix}\bigr)$ 
and $W_K = \bigl(\begin{smallmatrix} \bm{I}_D & \bm{0} & \bm{0} \\ \bm{0}^\top & -1 & 0 \end{smallmatrix}\bigr)$ 
(with $D\times D$ identity $\bm{I}_D$) give 
$\ket*{q_i} = (2\bm{r}_i, 1)^\top$ and 
$\ket*{k_j} = (\bm{r}_j, -r_j^2)^\top$, so 
$\braket*{q_i}{k_j} = 2\bm{r}_i\cdot\bm{r}_j - r_j^2 = r_i^2 - r_{ij}^2$; 
the $i$-only factor $\mathrm{e}^{\beta r_i^2}$ cancels in $Z_i$ to yield the Gaussian radial basis $Z_i^{-1}\mathrm{e}^{-\beta r_{ij}^2}$. 
On a periodic basis $\ket*{h_i} = (\cos\bm{k}\cdot\bm{r}_i, \sin\bm{k}\cdot\bm{r}_i)^\top$ at wavevector $\bm{k}$, $W_Q = W_K = \bm{I}_2$ directly yields $\cos[\bm{k}\cdot(\bm{r}_i-\bm{r}_j)]$
}
% Derivation pen and paper in 2026.C notebook (A4, yellow), p.16 or so. Could write up for SI?

Crucially, for strongly correlated pairs $(i,j)$, the Boltzmann factor concentrates $\mathcal{O}(1)$ weight on partner $j$ while silencing distant spectators (this is attention, $\Sigma_{ik}^{\text{Attn}} \to 0$ for $k \neq j$)), preserving an $\mathcal{O}(1)$ off-diagonal Jacobian $\bm{J}_{ij}^{\text{Attn}} = \pdv*{h_i^{(l+1)}}{\bm{r}_j} \propto \Sigma_{ij}^{\text{Attn}}[h^{(l)}]$ independent of system size $N$.

In continuum operator learning, Tsai et al.\cite{Tsai2019} and Cao\cite{Cao2021Galerkin} recently proved that self-attention acts as a discretised Fredholm integral operator whose kernel approximates a differential equation Green's function (with unnormalised linear attention forming a Petrov--Galerkin projection). 
Reformulating in explicitly quantum-mechanical notation, 
the residual skip connection ($\delta_{ij}$) acts as the bare one-body propagator $G_0$, 
the attention kernel $\Sigma_{ij}^{\text{Attn}}[h^{(l)}]$ acts as the pairwise self-energy functional $\Sigma[G_1]$, 
and the $L$-layer map dressing bare inputs into quasiparticles, $\ket*{h^{(L)}} = G_1\ket*{h^{(0)}}$, iterates Dyson's equation\cite{Dyson1949},
\begin{equation}
    G_1 = G_0 + G_0 \, \Sigma[G_1] \, G_1, \qquad
    \expect*{\hat{V}} = \frac{1}{2}\operatorname{Tr}\!\left(\Sigma[G_1]\,G_1\right).
    \label{eq:dyson_gm}
\end{equation}
We express this diagrammatically in Figure~\ref{fig:mattuck_dyson}.
Unrolling $L$ attention layers (with queries $\ket*{q_i^{(l)}}$ and keys $\ket*{k_j^{(l)}}$ recomputed from dressed states at each steps) generates the self-consistent Dyson series $G_1 = G_0 + G_0 \Sigma[G_1] G_0 + G_0 \Sigma[G_1] G_0 \Sigma[G_1] G_0 + \dots$, dressing single-particle coordinates into quasiparticles. 
% SERIOUSLY - isn't that astonishing?
Because physical Hamiltonians $\hat{H} = \sum_{i=1}^N \hat{h}(i) + \sum_{i<j}^N \hat{v}(i,j)$ contain at most pairwise ($\hat{v}(i,j)$) interactions, L\"owdin's 2-RDM theorem\cite{Lowdin1955-2RDM} and the Galitskii--Migdal relation\cite{Galitskii1958} for the interaction energy $\expect*{\hat{V}}$  guarantee that a dressed one-body propagator $G_1$ coupled to a two-body self-energy $\Sigma[G_1]$ suffices to determine exact energetics and two-body observables.
% Galtiskii-migdal:     \expect*{\hat{V}} = \frac{1}{2}\operatorname{Tr}\!\left(\Sigma[G_1]\,G_1\right).
% But don't think you need it here explicitly? 

This correspondence explains both the architectural split across atomistic machine learning and its computational scaling. 
By Kohn's nearsightedness principle\cite{Prodan2005}, screened, \textit{extensive} ground-state potential energy surfaces $E_0(\bm{R})$ depend only on short-range local environments, so finite-cutoff message passing followed by $\mathcal{O}(N)$ Hartree sum-pooling ($E = \sum_i E_i$, as in \textsc{MACE}\cite{Batatia2022} and other MLIPs) is physically complete. 

Conversely, nonlocal quantum observables depend on all $\binom{N}{2}$ ($N^2$) off-diagonal two-body propagators and relative fragment orientations, which are erased by mean-field sum-pooling but preserved by $\braket{q_i}{k_j}$. 
Both \textsc{PsiFormer}\cite{vonglehn2023selfattentionansatzabinitioquantum} in quantum Monte Carlo and \textsc{PET}\cite{Pozdnyakov2023PET} (albeit limited to local clusters) in interatomic potentials have empirically discoverd the physical rewards of self-attention. 
The clearest next frontier lies wherever models must reconstruct nonlocal quantum operators directly: two-index tight-binding and Kohn--Sham Hamiltonian matrix elements $H_{ij}(\bm{R})$
(such as \textsc{SchNOrb}\cite{Schtt2019SchNOrb}, \textsc{DeepH}\cite{Li2022DeepH} and \textsc{MACE-H}\cite{Qian2026MACE-H}),
, or the multi-state companion Hamiltonians $M_{mn}(\bm{R})$ that govern excited-state crossings in \textsc{X-MACE}\cite{Barrett2025}, where pairwise attention weights naturally mirror off-diagonal couplings.

The $\mathcal{O}(N^2)$ scaling of dense self-attention is thus not an architectural flaw, but the irreducible cost of unscreened pairwise interactions when nearsightedness fails. 
By further analogy with computational condensed matter, sparse and hierarchical attention mirror the development of fast-multipole methods.

\section{Results: Fractal strawberry}
% J: And it collapsed in on itself. 'cause you see, when you really put everything on a [fractal strawberry], it becomes this. The truth.
% E: What is the truth?
% J: Nothing matters. 

The insights in this paper stem from fitting the fractal determinant\cite{Kruger2008} of a hydrodynamic backflow wavefunction\cite{Feynman1956Backflow} to verify the expressive power of a continuous normalising flow (CNF) backflow model (\textsc{Spindrift}\cite{Frost2026Spindrift}). 
To model Fermionic quantum criticality, Kr\"uger and Zaanen\cite{Kruger2008} dressed free-Fermion plane waves of occupied momenta in a Slater determinant $\Psi_{\text{BF}}(\bm{R}) = \det[\exp(\mathrm{i}\bm{k}_m \cdot \tilde{\bm{r}}_n)]$, evaluated at quasiparticle backflow coordinates $\tilde{\bm{r}}_i$, displaced by an instantaneous two-body velocity field $\bm{v}_i(\bm{R})$,
\begin{align}
    \tilde{\bm{r}}_i &= \bm{r}_i + \bm{v}_i(\bm{R}), \notag \\*
    \bm{v}_i(\bm{R}) &= \sum_{j \neq i} \eta(r_{ij})(\bm{r}_i - \bm{r}_j), \notag \\*
    \eta(r_{ij}) &= \frac{a^3}{r_{ij}^3 + r_0^3},
    \label{eq:kruger_backflow}
\end{align}
where $r_{ij} = \lvert\bm{r}_i - \bm{r}_j\rvert$, $a$ controls the backflow strength, and $r_0$ sets the short-range length scale. 
Beyond a critical coupling strength in periodic systems, the nodal surface $\Psi_{\text{BF}}(\bm{R})=0$ undergoes a topological transition from smoothly varying (Fermi-liquid like) into a scale-invariant fractal, driven by what Kr\"uger and Zaanen identified diagrammatically as infinite-order vertex corrections\cite{Kruger2008}. 
This wavefunction was a putative suggestion based on scaling arguments, with no construction of the underlying Hamiltonian offered. 
The form of the backflow comes from hydrodynamic flow arguments in the original setting of liquid helium\cite{Feynman1956Backflow}. 

Although this fractal nodal surface embodies infinite-order many-body criticality, the instantaneous backflow velocity field has the exact pairwise Dyson structure of Eqn.~\eqref{eq:master_dyson}, with self-energy kernel $\Sigma_{ij}^{\text{BF}} = \eta(r_{ij})$ weighting the pairwise relative-coordinate message $\ket*{v_{ij}} = \bm{r}_i - \bm{r}_j$. 
In \textsc{Spindrift}\cite{Frost2026Spindrift}, we evaluate an open-boundary thermal Slater determinant ($N=6$ spin-polarised fermions in $D=2$) after calculating quasi-position (backflow) from integrating a single-layer ($L=1$) neural velocity field over pseudo-time $s \in [0, 1]$. 
In order to induce fractal phase space folding with open boundaries, we integrate with only two coarse RK4 steps ($\Delta s = 0.5$).

\begin{figure}
    \centering
    \includegraphics[width=1.0\linewidth]{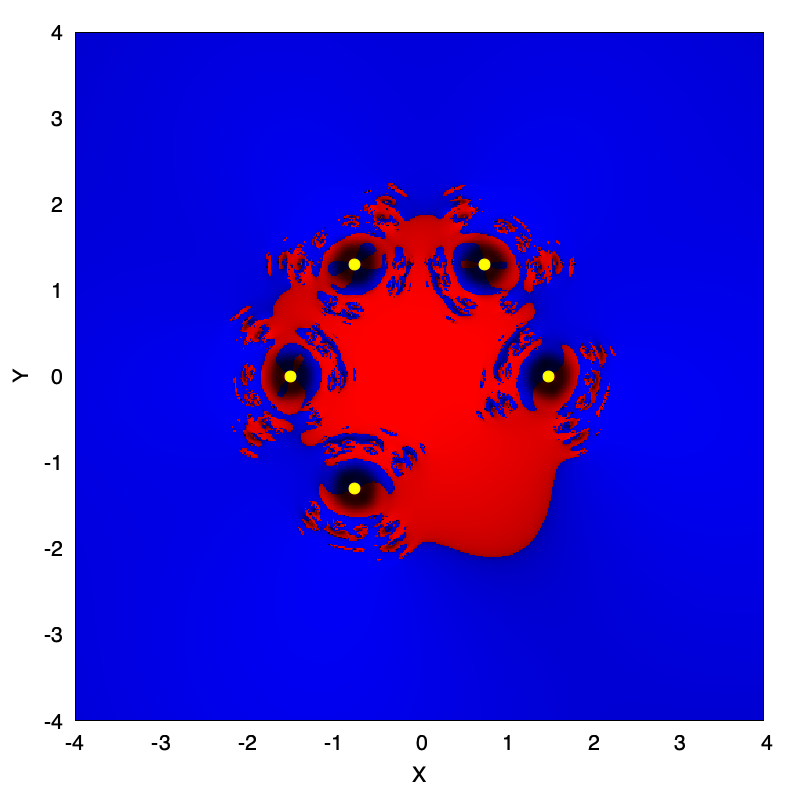}
    \caption{The reference fractal backflow determinant which we used for fitting, $r_0 = 0.2$ and $a=1$ (strongly fractal). Visualised for $N=6$ by holding 5 particles stationary (yellow dots) on a ring of radius 1.5 and then using the sixth particle as a probe (red/blue: determinant sign; brightness: $|\!\det\!|^{0.07}$, nodes in black). 
    In this open-boundary model, the nested fractal pockets emerge from phase-space folding across the two intentionally coarse RK4 steps ($\Delta s = 0.5$) that integrate the singular $1/r^3$ velocity field into quasi-positions.
    }
    \label{fig:kruger_backflow_fractal}
\end{figure}

\begin{figure*}
        \centering
        \includegraphics[width=1.0\linewidth]{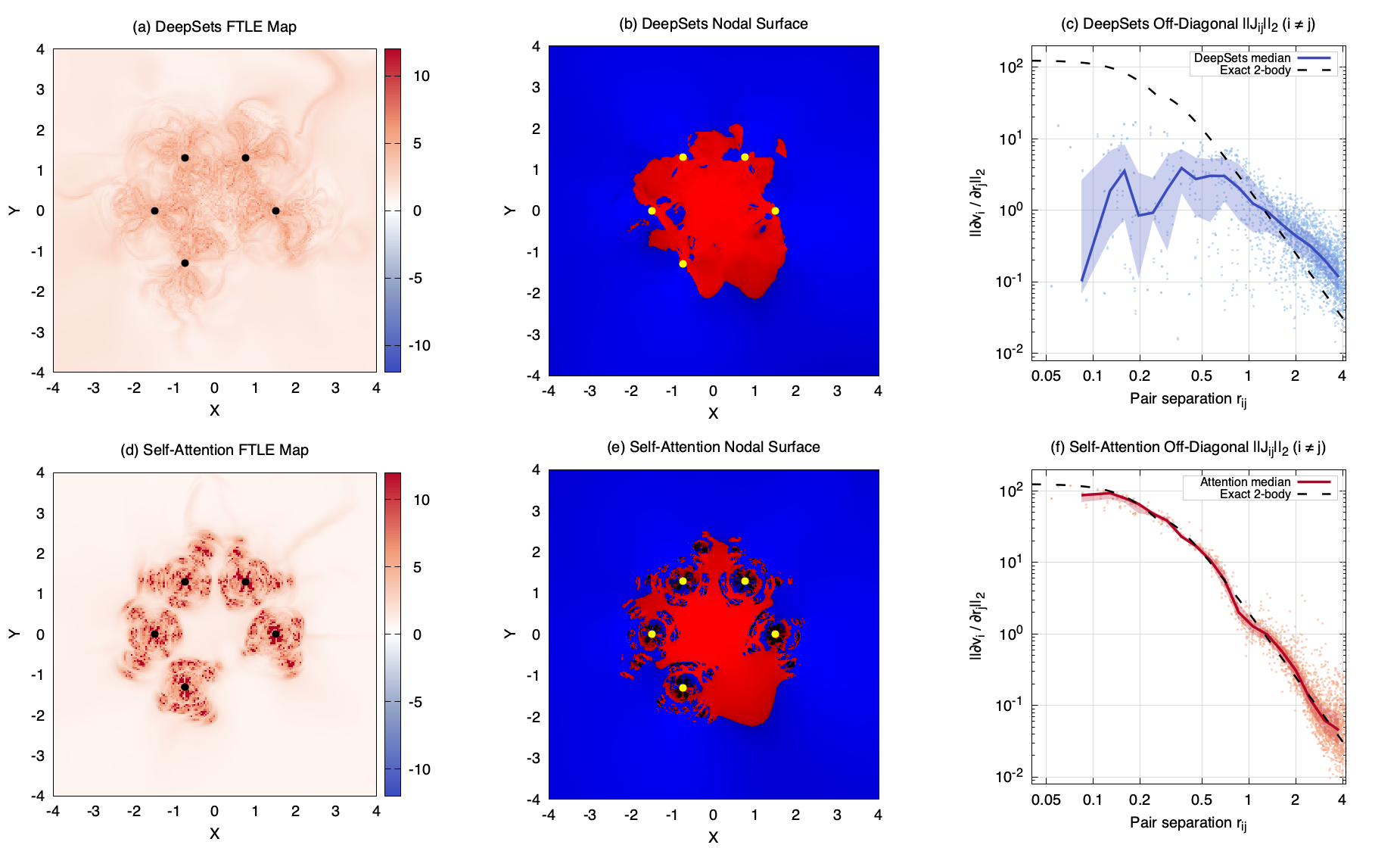}
        \caption{Width-matched DeepSets (top) and self-attention (bottom) fits to the fractal backflow target
        (Fig.~\ref{fig:kruger_backflow_fractal}), trained by $L_1$ regression on the instantaneous velocity field of
	Eqn.~\eqref{eq:kruger_backflow}, using the same coarse two-step RK4 integrator.
    \textbf{(a,\,d)} Finite-time Lyapunov exponent (FTLE) maps $\ln\sigma_{\max}(\nabla_{\bm{R}}\tilde{\bm{R}})$, where $\sigma_{\max}$ denotes the maximum singular value of the flow-map Jacobian $\nabla_{\bm{R}}\tilde{\bm{R}}$: DeepSets smears sensitivity (albeit rather prettily) across the trap, whereas self-attention resolves dipolar rosettes around the stationary particles (black dots).
	\textbf{(b,\,e)} Slater determinants: DeepSets collapses the nodal surface into a smooth mean-field envelope, whilst self-attention recovers the nested fractal pockets.
    \textbf{(c,\,f)} Off-diagonal two-body velocity Jacobian spectral norm $\lVert\partial\bm{v}_i/\partial\bm{r}_j\rVert_2 = \sigma_{\max}(\partial\bm{v}_i/\partial\bm{r}_j)$ (the largest singular value of each $2\times 2$ pair block $i\neq j$ of $\pdv*{\bm{v}}{\bm{R}}$) versus pair separation $r_{ij}$ across $3{,}600$ random pairs (points are raw samples; solid line and shaded band show median and
	interquartile range; black dashed line shows the exact analytic derivative): DeepSets suffers $\mathcal{O}(1/N)$ saturation at short range alongside spurious crosstalk at long range, whereas self-attention tracks the exact pairwise kernel
	over four decades. 
%    From the individual datapoints this crosstalk was clearly pathological: close approach of two particles would then induce a shift very far away.
%    
Both velocity networks share identical encoders and decoders (hidden dimension $d_{\text{model}} = 128$, \textsc{GELU} activations, zero-initialised output layer; $50{,}690$ \texttt{Float64} parameters). 
DeepSets then uses parameter-free mean-pooling, whereas a $H=4$-head $d_k = 32$ attention block adds $65{,}536$ parameters for a total of $116{,}226$.
Both models were trained by online stochastic sampling for $10{,}000$ minibatch steps (batch size $512$), with the $L_1$ loss saturating for DeepSets at $\approx 1.0$ after $200$ steps, while self-attention descended smoothly to $0.17$.}
\label{fig:fitted_backflow}
\end{figure*}

When we attempted to learn this backflow mapping using an equivariant DeepSets architecture (Eqn.~\eqref{eq:deepsets}), the $\mathcal{O}(1/N)$ mean-field sum washed out the pairwise kernel, killed the local shear ridges in the finite-time Lyapunov exponent (FTLE) map (Fig.~\ref{fig:fitted_backflow}a), and collapsed the fractal structure (Fig.~\ref{fig:kruger_backflow_fractal}) into a smooth, featureless envelope---a ``squashed strawberry'' (Fig.~\ref{fig:fitted_backflow}b). 
The model was simply not expressive enough to hold the inter-particle correlation which generated the fractal structure. 
Replacing DeepSets pooling with a \textsc{PsiFormer}-inspired self-attention layer ($\sim 20$ lines of Julia code) restores the $\mathcal{O}(1)$ nonlocal Dyson kernel $\Sigma_{ij}$, recovering both the dipolar FTLE rosettes (Fig.~\ref{fig:fitted_backflow}d) and the self-similar fractal nodal pockets (Fig.~\ref{fig:fitted_backflow}e).

Differentiating the fitted velocity networks $\bm{v}_i(\bm{R})$  directly provides our key quantitative test of Eqn.~\eqref{eq:jacobian_ds} (Fig.~\ref{fig:fitted_backflow}c,\,f). 
%These are the key numeric results of this letter. 
As the probe particle approaches a stationary ring particle ($r_{ij} = 0.05 < r_0$), the exact off-diagonal velocity Jacobian norm reaches $\lVert\partial\bm{v}_1/\partial\bm{r}_6\rVert_2 = 123$: self-attention reproduces this within $1.6\%$ ($125$), whereas DeepSets saturates at $\sim 8$ from $r_{ij} < 0.8$. . 
Simultaneously, for the diametrically opposite spectator particle ($r_{ij} = 3.05$), the true $r_{ij}^{-3}$ coupling decays to $0.070$ (the self-attention fit is $0.021$), yet DeepSets predicts $0.325$, producing the vertical spray of long-range outliers in Fig.~\ref{fig:fitted_backflow}c: inflating $\pdv*{v_j}{\bm{r}_j}$  to chase a short-range collision broadcasts a global shift through $\overline{\Sigma}$ to every particle in the system. 

From a renormalisation group (RG) perspective, Kr\"uger and Zaanen's fractal nodal surface is the geometric signature of a quantum critical fixed point, where scale-invariant vertex corrections fold phase space across all length scales\cite{Kruger2008}. 
Under DeepSets sum-pooling, the $\mathcal{O}(1/N)$ Jacobian collapse acts as a hard ultraviolet cutoff that amputates short-range gradients (Fig.~\ref{fig:fitted_backflow}c), trapping the flow at a trivial mean-field fixed point: the smooth ``squashed strawberry'' (Fig.~\ref{fig:fitted_backflow}b). 
Self-attention preserves the power-law two-body kernel across four decades of separation (Fig.~\ref{fig:fitted_backflow}f), allowing repeated composition of the flow map to sustain criticality and generate the self-similar fractal attractor (Fig.~\ref{fig:fitted_backflow}e).

\section{Conclusions}

Taking this Dyson equation correspondence seriously explains specific Transformer design choices. 
First, the exponential Boltzmann factor in Eqn.~\eqref{eq:attention_dyson} is essential: whereas $\mathcal{O}(N)$ ``linear attention'' ($\Sigma_{ij} \propto \braket{q_i}{k_j}$) reduces to a separable low-rank kernel (i.e. a finite-order perturbation theory), expanding $\exp(\beta\braket{q_i}{k_j})$ generates all powers of the two-body vertex, providing a full-rank kernel. 
Second, had this kernel been evaluated solely on bare input coordinates, stacking $L$ layers would merely unroll a bare Born perturbation series $G_0 + G_0\Sigma[G_0]G_0 + \dots$. 
Because a Transformer instead recomputes queries and keys from the already-dressed states at every layer, it feeds the dressed propagator back inside the self-energy functional, $\Sigma[h^{(l)}]$[. 
Diagrammatically, network depth executes a self-consistent skeleton expansion (summing polarisation rings and vertex dressings in a discrete functional RG flow).
Multi-head attention decomposes this self-energy action into $H$ parallel interaction channels, $\sum_{h=1}^H \Sigma_{ij}^{(h)}[h^{(l)}]\ket*{v_j^{(l,h)}}$ (where $\ket*{v_j^{(l,h)}}$ absorbs each head's output projection, as seen in \textsc{PsiFormer} on the benzene dimer\cite{vonglehn2023selfattentionansatzabinitioquantum}, where separate heads spontaneously partition short-range intramolecular exchange from long-range intermolecular dispersion).

This perturbative structure likewise explains why residual skip connections and zero-initialised output layers are often required to train deep attention networks. 
Switching on a strong self-energy $\Sigma$ instantaneously drives the Dyson series $(G_0^{-1} - \Sigma)^{-1}$ across poles and branch cuts; bare states deform smoothly into dressed quasiparticles only when the interaction is turned on adiabatically from the non-interacting propagator $G_0$. 
Deep Transformers and continuous normalising flows rediscover this exact safeguard. The skip path supplies the bare propagator $G_0 = \bm{I}$, while zero-initialising the output projection pins the initial model at the non-interacting limit, $\Sigma = 0$, allowing gradient descent to dress coordinates adiabatically without crossing singularities. 
(Replacing spatial indices with time steps immediately extends this picture to open quantum systems, where causal self-attention acts as a non-Markovian Nakajima--Zwanzig memory kernel.)

Equivariant sum-pooling and self-attention are not just two competing heuristics; they mirror the two foundational approximations that have structured condensed-matter physics for a century.
Sum-pooling is a mean-field (Hartree) approximation. 
Where the nearsightedness principle holds (as in ground-state interatomic potentials) summing local one-body energies is both efficient and physically complete. 
Where nearsightedness breaks down (such as correlated wavefunction determinants, or long-range excited states), sum-pooling suffers an $\mathcal{O}(1/N)$ Jacobian collapse that washes out off-diagonal structure into a featureless mean-field.

Self-attention fixes this by replacing a one-body sum with a real-space Dyson iteration. 
Its softmax-normalised inner product constructs a state-dependent, two-body self-energy $\Sigma_{ij}$ on the fly, preserving the $\mathcal{O}(1)$ pairwise sensitivity required to reconstruct scale-invariant fractal attractors and nonlocal excitonic couplings. 
Because physical Hamiltonians are built from pairwise interactions, L\"owdin's 2-RDM theorem and the Galitskii--Migdal relation guarantee that a dressed one-body propagator coupled to a two-body self-energy is formally complete for many-body energetics.
The rise of the Transformer is therefore no accident; it rediscovers the minimal complete representation demanded by pairwise interactions. 
Mean-field pooling suffices for local potentials, but for correlated matter, attention is all we ever needed.

\begin{acknowledgments}
I thank Christoph Ortner, Matthew Foulkes and Frank Kr\"uger for fruitful discussion that led directly to the realisations described in this letter. 
J.M.F.\ is supported by a Royal Society University Research Fellowship (URF-R1-191292). 
Julia codes implementing \textsc{Spindrift} and the fractal backflow benchmarks are available on GitHub as part of \textsc{Halcyon.jl}\cite{Halcyon}. 
I gratefully acknowledge the use of the Imperial College Research Computing Service\cite{HPC}.
\end{acknowledgments}

\bibliography{2026-11-FractalStrawberry}

\end{document}